\pdfoutput=1
\documentclass[aps,pre,twocolumn]{revtex4}

\usepackage{graphicx}
\usepackage{amssymb,amsfonts,amsmath}
\usepackage{hyperref}
\usepackage{longtable}
\setlongtables

\begin{document}

\title{Excitations are localized and relaxation is hierarchical in glass-forming liquids}

\author{Aaron S. Keys} 
\thanks{These authors contributed equally to this work.}
\affiliation{Department of Chemistry, University of California, Berkeley CA, 94720}
\affiliation{Lawrence Berkeley National Laboratory, Berkeley CA, 94720}
\affiliation{Department of Chemical Engineering, University of Michigan, Ann Arbor MI, 48109-2136}
\author{Lester O. Hedges}
\thanks{These authors contributed equally to this work.}
\affiliation{Department of Chemistry, University of California, Berkeley CA, 94720}
\affiliation{Lawrence Berkeley National Laboratory, Berkeley CA, 94720}
\author{Juan P. Garrahan}
\affiliation{School of Physics and Astronomy, University of Nottingham, Nottingham, NG7 2RD, United Kingdom}
\author{Sharon C. Glotzer}
\affiliation{Department of Chemical Engineering, University of Michigan, Ann Arbor MI, 48109-2136}
\affiliation{Department of Materials Science Engineering, University of Michigan, Ann Arbor MI, 48109-2136}
\author{David Chandler}
\email[Corresponding author.  E-mail: ]{chandler@berkeley.edu}
\affiliation{Department of Chemistry, University of California, Berkeley CA, 94720}

\date{\today}

\begin{abstract}
For several atomistic models of glass formers, at conditions below their glassy dynamics onset temperatures, ${T_\mathrm{o}}$, we use importance sampling of trajectory space to study the structure, statistics and dynamics of excitations responsible for structural relaxation.
Excitations are detected in terms of persistent particle displacements of length $a$.  At supercooled conditions, for $a$ of the order of or smaller than a particle diameter, we find that excitations are associated with correlated particle motions that are sparse and localized, occupying a volume with an average radius that is temperature independent and no larger than a few particle diameters.  We show that the statistics and dynamics of these excitations are facilitated and hierarchical.  Excitation energy scales grow logarithmically with $a$.  Excitations at one point in space facilitate the birth and death of excitations at neighboring locations, and space-time excitation structures are microcosms of heterogeneous dynamics at larger scales.  This nature of dynamics becomes increasingly dominant as temperature $T$ is lowered.  We show that slowing of dynamics upon decreasing temperature below $T_\mathrm{o}$ is the result of a decreasing concentration of excitations and concomitant growing hierarchical length scales, and further that the structural relaxation time $\tau$ follows the parabolic law, $\log(\tau / \tau_\mathrm{o}) = J^2(1/T - 1/T_\mathrm{o})^2$, for $T<T_\mathrm{o}$, where $J$, $\tau_\mathrm{o}$ and $T_\mathrm{o}$ can be predicted quantitatively from dynamics at short time scales. Particle motion is facilitated and directional, and we show this becomes more apparent with decreasing $T$. We show that stringlike motion is a natural consequence of facilitated, hierarchical dynamics.
\end{abstract}

\maketitle

\section{Introduction}

The behaviors of supercooled glass-forming liquids manifest complex correlated particle dynamics.  Signature behaviors{, which appear below an onset crossover temperature, $T_\mathrm{o}$,} include super-Arrhenius growth of relaxation times with lowering temperature, stretched exponential time-correlation functions, and transport decoupling.  Palmer et al.~\cite{palmer1984models} suggest that these behaviors follow from hierarchical dynamics in which excitations on one scale facilitate dynamics of neighboring excitations~\cite{glarum1960dielectric}  thereby creating excitations on larger scales.  This idea is encoded by a class of dynamical models, so-called kinetically constrained models (KCMs)~\cite{fredrickson1984kinetic, ritort2003glassy}.  Results from one such model agree with experimental observations of signature dynamical behaviors~\cite{garrahan2003coarse,elmatad2009corresponding,chandler2010dynamics}.  KCMs presume that excitations are localized and free of any significant static inter-excitation correlation.  All large scale effects are due solely to the nature of excitation dynamics, which is facilitated and directional.  This paper presents our discovery that for several different atomistic models, these features are emergent properties of underlying Newtonian dynamics.

\subsection{Molecular dynamics simulations demonstrate facilitated and hierarchical dynamics.} 

For each of five atomistic models at several different densities and temperatures, we ask the question:  How does an atom move a distance $a$ between relatively long lived neighboring positions?  We answer this question by augmenting extensive molecular dynamics simulations with methods of transition path sampling~\cite{bolhuis2002transition}.  In particular, molecular reorganization in a supercooled liquid is a rare event, and transition path sampling can harvest unbiased ensembles of trajectories exhibiting such events.  By applying this approach, we find several important results.

First, for $a$ smaller than or not much larger than a particle diameter, {particle displacements that stick to a new position for a significant period of time are associated with correlated displacements of only a handful of neighboring particles.  These persistent changes in particle position, which serve as indicators of what we call ``excitations,'' }are closely related to the micro-strings~\cite{gebremichael2004particle} found in earlier computer modeling studies of heterogeneous dynamics in glass forming liquids.   At supercooled conditions, i.e., at temperatures $T$ below the onset temperature, $T_\mathrm{o}$, their occurrence is sparse.  They arise in localized regions of relatively high mobility whose size is largely independent of temperature, and whose spatial distribution is that of a dilute gas. 

Second, for a given displacement length $a$, we find that the equilibrium concentration of excitations, $c_a$, has a Boltzmann temperature dependence
\begin{equation}
\label{eqn:cvsT}
c_a \propto \exp\left[-J_a \left(1/T - 1/T_\mathrm{o}\right)\right], \quad T<T_\mathrm{o},
\end{equation}
where $J_a$ grows logarithmically with $a$, {i.e., 
\begin{equation}
\label{eqn:Jloga}
J_a - J_{a'} = \gamma \, J_\sigma \, \ln \left( a/a' \right)\,.
\end{equation}
Here, $\sigma$ is the space-filling diameter of a particle, the values of $\gamma$ and $J_\sigma$ are material-dependent, and $\gamma$ is of order unity. } 

Third, we find that observed super-Arrhenius temperature variation of transport properties follows directly from this logarithmic scaling of $J_a$.  In particular, as the average distance between excitations grows with lowering $T$, the energy scale for relaxation also grows with lowering $T$.  Specifically, from Eq.~(\ref{eqn:Jloga}), we argue that the structural relaxation time, $\tau$, obeys
\begin{equation}
\label{eqn:tauvsT}
\tau = \tau_\mathrm{o} \exp \left[ J^2 \left(1/T - 1/T_\mathrm{o}\right)^2 \right], \quad T< T_\mathrm{o}\,,
\end{equation}
with $J = \sqrt{\gamma /d_\mathrm{f}\,}\,J_\sigma$, where $d_\mathrm{f}$ is the fractal dimensionality of heterogeneous dynamics.  The reference time, $\tau_\mathrm{o}$, depends little upon temperature, and it is of the order of the time to reorganize a local arrangement of particles in the presence of an excitation, which in turn is of the order of structural relaxation times of the liquid at temperatures above $T_\mathrm{o}$.  We show that Eq.~(\ref{eqn:tauvsT}) holds quantitatively with $d_\mathrm{f}/d \approx 0.9$ and 0.8 for physical dimensions $d=2$ and $3$, respectively.

\subsection{Results demonstrate the perspective of kinetically constrained models.}

Equation~(\ref{eqn:tauvsT}) is the temperature dependent form for $\tau$ that successfully collapses disparate transport properties of all fragile glass formers~\cite{elmatad2009corresponding, elmatad2010corresponding}.  The logarithmic scaling from which it results, Eq. \ref{eqn:Jloga}, is that of the $d=1$ East model~\cite{eastmodel} and higher dimensional generalizations \cite{garrahan2003coarse,Berthier:2005hb}.  Here, by providing a microscopic recipe for detecting excitations, computing $c_a$ and predicting $\tau$, we validate East-like KCMs as good models for the dynamics of atomistic glass formers.

These technical and quantitative advances seem especially significant in light of the implied physical picture.  In particular, we establish herein that the principal growing length that governs the slowing of dynamics in a structural glass former is the mean distance separating excitations, $\ell_a \propto c_a^{-1/d_\mathrm{f}}$.  Relaxation requires the correlated dynamics of neighboring excitations, and the greater their separation, the longer it takes to coordinate their motions.   It is a picture that appears significantly different than an often invoked structural perspective originating with Adam and Gibbs~\cite{adam1965temperature,Cavagna:2009rt}.  That alternative imagines an underlying mosaic of ordered domains, so-called ``cooperative rearranging regions.''  Relaxation follows from reorganizing these regions~\cite{xia2000fragilities,biroli2009random, berthier2011theoretical}, and dynamics slows because mosaic domains grow.  Our findings do not preclude a picture based upon growing cooperative rearranging regions, but a link between it and the localized excitations that we document would need to relate $\ell_a$  to the size of mosaic elements, and whether an operational definition of the latter can accomplish this task is unclear.

\section{Methods}

The abundance of results collected in this paper are unprecedented by the current standards of this field.  These results establish the behaviors of many distinct large systems, and demonstrate the commonality of these behaviors over a broad range of time scales.  {Even so, further studies could provide additional documentation and perhaps refinements of the conclusions we are able to draw from the results presented herein.}

\subsection{We study the molecular dynamics of five distinct simple liquid mixtures.}

The atomistic models we have simulated are two-component mixtures for which particles of type $\alpha$ interact with those of type $\gamma$  with pair potentials, $u_{\alpha \gamma}(r)$, where $r$ is the separation of the pair.  These potentials are shifted and truncated Lennard-Jones potentials,
\begin{equation*}
u_{\alpha \gamma}(r) 
= \left\{
\begin{array}{lr}
u^{\mathrm{(LJ)}}(r\,; \sigma_{\alpha \gamma}, \epsilon_{\alpha \gamma}) - 
& \\
\quad u^{\mathrm{(LJ)}}(r_{\alpha \gamma}^{(\mathrm{c})}\,;  \sigma_{\alpha \gamma}, \epsilon_{\alpha \gamma}) \, ,
&
r\leqslant r_{\alpha \gamma}^{(\mathrm{c})} \\
& \\
0 \, ,
&
\quad r>r_{\alpha \gamma}^{(\mathrm{c})} 
\\
\end{array}
\right.
\end{equation*}
where $u^{(\mathrm{LJ})}(r\,; \sigma, \epsilon) = 4 \epsilon [(\sigma/r)^{12} - (\sigma/r)^{6}]$.
For each, we use $N=10,000$ particles in total, with $f N $ and $(1-f)N$ being the number of A-particles and B-particles, respectively.  The different models are distinguished by different choices of length and energy parameters, of mixing fraction $f$, and of particle masses $m_\mathrm{A}$ and $m_\mathrm{B}$.  For notational ease, we use dimensionless quantities throughout, where Boltzmann's constant is unity, and the unit of time is $(\epsilon / m \sigma^2)^{-1/2}$, where $\epsilon = \epsilon_\mathrm{AA}$, $m = m_\mathrm{A}$, and $\sigma = \sigma_\mathrm{AA}$ are the units of energy, mass, and length, respectively.  
\begin{table*}
\caption{\label{table:param} Parameters for model systems used in this study.$^a$ }
\begin{tabular}{ l | c c c c c c c c c c c c c }
\hline\hline   \  
 \ model & $d$ & $f$ &  $\sigma_\mathrm{AA}$ & $ \sigma_\mathrm{AB} $ & $ \sigma_\mathrm{BB}$ &  $ \epsilon_\mathrm{AA} $ &  $ \epsilon_\mathrm{AB} $ & $ \epsilon_\mathrm{BB} $ & $ r_\mathrm{AA}^{(\mathrm{c})} $ &$  r_\mathrm{AB}^{(\mathrm{c})} $ &  $ r_\mathrm{BB}^{(\mathrm{c})} $ & $ m_\mathrm{A} $ & $ m_\mathrm{B} $ \\
 \hline
KA$^b$ & 3 & 0.8 & 1.0 & 0.8 & 0.88 & 1.0 & 1.5 & 0.5 & 2.5& 2.5 & 2.5 & 1.0 & 1.0 \\
W$^c$ & 3 & 0.5 & 1.0 & 11/12 & 5/6 & 1.0 & 1.0 & 1.0 & 2.5 & 2.5 & 2.5 & 2.0 & 1.0 \\
WCA3D$^d$ & 3& 0.5 & 1.0 & 11/12 & 5/6 & 1.0 & 1.0 & 1.0 & $2^{1/6}$ &  $(2^{1/6}11/6)$ &  $(2^{1/6}5/6)$ & 2.0& 1.0 \\
2D-50:50$^e$ & 2 & 0.5 & 1.0 & 1.2 & 1.4 & 1.0 & 1.0 & 1.0 & $2^{1/6}$ &  $(2^{1/6}1.2)$ &  $(2^{1/6}1.4)$ & 1.0 & 1.0 \\
2D-68:32$^f$ & 2 & 0.32 & 1.0 & 1.1 & 1.4 & 1.0 & 1.0 & 1.0 & $2^{1/6}$ &  $(2^{1/6}1.1$ )&  $(2^{1/6}1.4)$ & 1.0 & 1.0 \\
\end{tabular}
\begin{minipage}[t]{0.85\textwidth}
\footnotetext{$^a$Dimensionless units employed.  See text.} 
\footnotetext{$^b$Kob-Andersen model from Ref.~\cite{KA}.} 
\footnotetext{$^c$W\"{a}hnstrom model from Ref.~\cite{wahn}.}
\footnotetext{$^d$Weeks-Chandler-Andersen repulsive force mixture from Ref.~\cite{hedges2007decoupling}.}
\footnotetext{$^e$WCA potentials with the same particle-size ratios and mixing rules as the inverse power potentials of Ref.~\cite{hurley1995kinetic}.}
\footnotetext{$^f$WCA potentials with the same particle-size ratios and mixing rules as the inverse power potentials of Ref.~\cite{candelier2010spatiotemporal}.}
\end{minipage}
\end{table*}

The first three models detailed in Table~\ref{table:param} have been developed and studied by others before us.  The last two are modifications of models examined by Harrowell and co-workers~\cite{hurley1995kinetic, candelier2010spatiotemporal}.  In particular, we replace the inverse-twelve power potential of Refs.~\cite{hurley1995kinetic}~and~\cite{candelier2010spatiotemporal} with Weeks-Chandler-Andersen~\cite{weeks1971role} (WCA) repulsive potentials.  

By exploring the dynamics of these models for larger system sizes and longer times than have been considered before, we find that both variants of the 2D-50:50 system appear to coarsen and freeze.  As such, we use those systems for illustrative qualitative studies over timescales smaller than coarsening times.  For quantitative behavior of reversible transport and related properties, we use the other four models, none of which exhibit coarsening at the conditions we consider.  Our molecular dynamics for those systems is carried out with $N$, $V$ and $T$ fixed, using the HOOMD-Blue simulation code~\cite{anderson2008general} on graphics processors.

\subsection{Excitations are manifested by particle displacements that persist for significant periods of time.}

While glassy systems tend to be locally rigid or jammed, some locations are atypically less rigid and provide opportunities for structural reorganization.  These locations coincide with defects or excitations in an otherwise jammed material.  Their presence can be inferred or detected by observing non-trivial particle displacements associated with transitions between relatively long-lived configurations.  Because these displacements are reliable recorders of excitations, we often treat them as synonymous.  Nevertheless, the two are distinct. Displacements refer to dynamics in small parts of trajectories, while excitations refer to underlying configurations or micro-states.  We further comment on this point later in this paper.

Non-trivial displacements~-- the recorders of excitations~-- are different than fleeting vibrations.  Inherent structures~\cite{stillinger1984packing, heuer2008exploring} help distinguish the former from the latter.  The inherent structure of a configuration of an $N$-particle system is obtained by steepest descent to the nearest minimum in the potential energy landscape~\cite{bitzek2006structural}.  The inherent structure evolves as dynamics progresses, but most intra-basin vibrations are not apparent in that evolution.  Excitations are present when the dynamics produces significant changes in inherent structure. The same effect of removing relatively high-frequency vibrations can be obtained by coarse graining coordinates over a small increment of time.  We have generally chosen this latter procedure.  But we have checked in a few cases that our principal results for large enough length scales and time scales are unaffected by this choice, provided the coarse-graining time is of the order of the typical period for atomic length scale intrabasin vibrations and this time is of the order of, but generally smaller than, the typical instanton time $\Delta t$, defined below.  Where $\mathbf{r}_i(t)$ denotes the instantaneous position of the $i$th particle in the system at time $t$, we use ${\bar{\mathbf{r}}}_i(t)$ to denote its corresponding position in the time-coarse-grained structure, i.e.,
\begin{equation}
\bar{\textbf{r}}_i(t) = \frac{1}{\delta t} \int_0^{\delta t} dt' \ \textbf{r}(t+t'),
\end{equation}
where $\delta t$ is the coarse-graining time.  On the occasions where we refer to inherent structure coordinates, we use $\mathbf{r}_i^{\mathrm{IS}}(t)$ to denote the position of particle $i$ in the inherent structure at time $t$. 

\begin{figure*}
\label{fig:fig1}
\centerline{\includegraphics[width=1\textwidth]{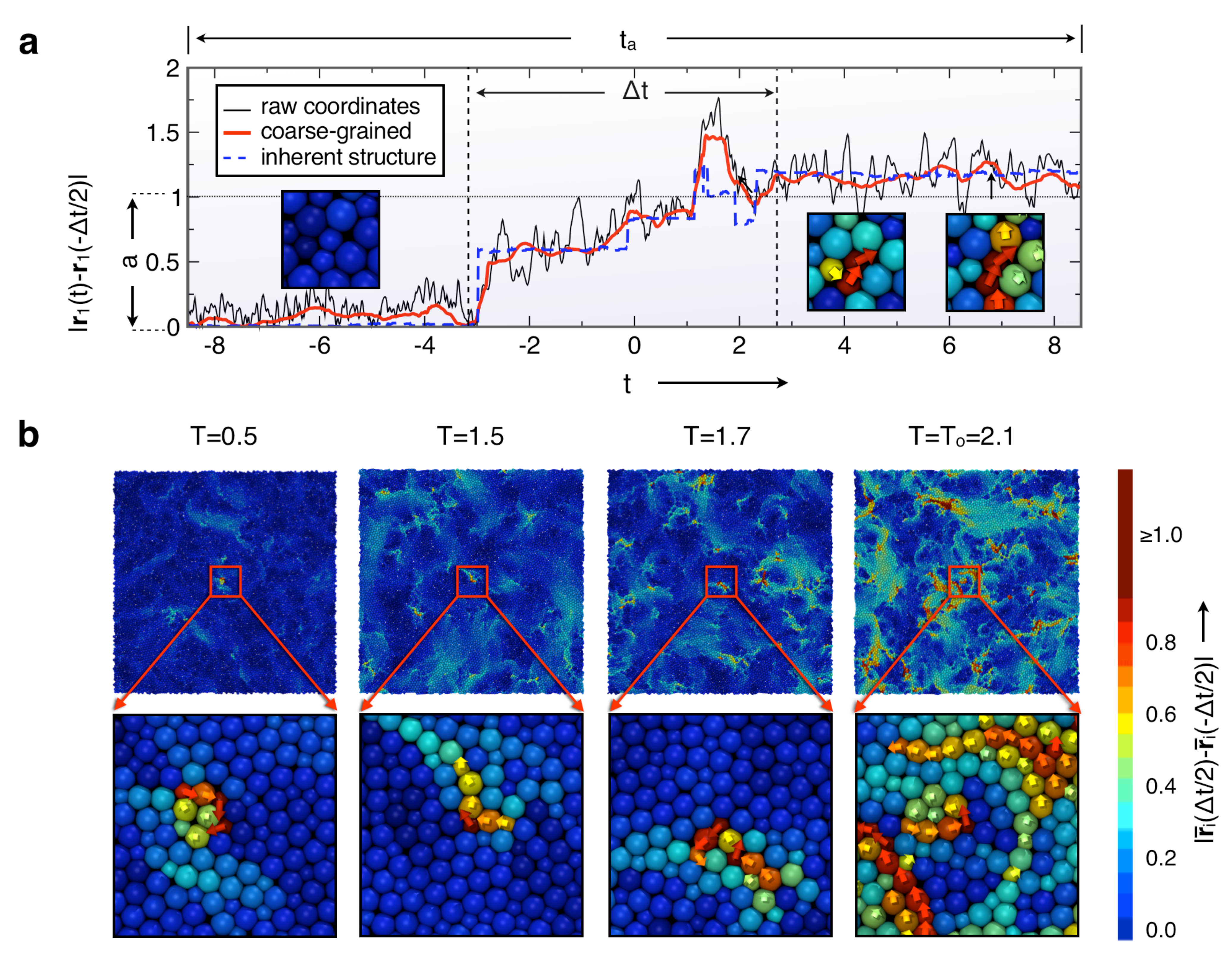}}
\caption{\label{fig:fig1} Excitations can be identified from dynamical processes localized in space and time.  Here, identifications are illustrated with rendered representative trajectories for the 2D-68:32 system (models in Table~\ref{table:param}). The unit of length is $\sigma$ and the unit of time is $(\epsilon / m \sigma^2)^{1/2}$; i.e, see text.  ({\rm a}) The upper panel shows the displacement for the unprocessed coordinate of a tagged particle at time $t$ together with that displacement in the inherent structure coordinate and in the time coarse-grained structure, i.e., $|\bar{\mathbf{r}}_1(t) - \bar{\mathbf{r}}_1(-\Delta t / 2)|$.  Here $\Delta t$ is the instanton time and $t_a$ is the plateau time.  The averaging window used is $\delta t = 0.6$.  The snapshots are drawn to show the space-filling sizes of particles and are colored to indicate the distance of a particle from its position at $t=-\Delta t / 2$.  The arrows indicate the direction of that motion.  ({\rm b}) The lower panels show similarly rendered snapshots for the full 10,000-particle system over one instanton time for a tagged particle at the center of the box.  Enlargements show detailed structures of the excitation dynamics, where arrows indicate the direction of the color-labeled displacements.  Blue indicates relative immobility while red indicates relative mobility.  The depictions are for particle density $\rho = 0.75$ at four different temperatures (as indicated).  The glassy dynamics onset temperature for this system at this density is $T_\textrm{o} \approx 2.1 $.}
\end{figure*}

Figure~\ref{fig:fig1} renders part of a trajectory from one of the models we have considered.  It illustrates typical behavior of these processed coordinates.  While inherent structures change discontinuously as the system moves between potential energy basins, the raw and coarse-grained coordinates are continuous.  The particular trajectory illustrated in Fig.~\ref{fig:fig1}(a) shows the appearance of the type of motion with which we detect an excitation.  It is characterized by three periods:  During the first, the inherent structure of a tagged particle remains near its initial point; following this sojourn, the system undergoes a rapid transition that lasts up to a time $\Delta t$, termed the ``instanton'' time, after which the tagged particle is found in a distinctly different position for another relatively quiescent period.  

This behavior suggests using the following functional of path as the indicator that at time $t$ particle $i$ is associated with an excitation with displacement length $a$:
\begin{equation}
\label{indicator}
h_i(t, t_a; a) = \prod_{t' =\, t_a/2 - \Delta t  }^{ t_a / 2} \theta \left(|\bar{\mathbf{r}}_i(t + t') - \bar{\mathbf{r}}_i(t-t')| - a \right) \,.
\end{equation}
Here, $\theta(x)=1$ or 0 for $x \geq 0$ or $<0$, respectively. The products are over time steps of a trajectory that extends for a time $t_a > \Delta t$.  This length of time is at least as long as the time for trajectories to commit to one basin or the other while traversing a transition state region.  In the parlance of rare-event dynamics, $t_a$  is a plateau or commitment time~\cite{chandler1978statistical,hanggi1990reaction} -- a time for which the equilibrium average of $h_i(t, t_a; a)$ grows linearly with $t_a$.  It is typically 3 or 4 times the mean instanton time for that displacement length $a$, $\langle \Delta t \rangle_a$.  As such, the indicator given by Eq.~\ref{indicator}  discards trajectories that exhibit displacements with length scale $a$ if they fail to persist for at least as long as the typical instanton time.  The instanton time is the shortest time separating the initial and final sojourns. It is a functional of path, and it varies from one trajectory to another.  

Fig.~\ref{fig:fig1}(b) illustrates the spatial arrangements of these excitations.  In particular, at supercooled temperatures, for reasonable choices of $a$, the pictures give the impression that the mean excitation concentration is small.  As temperature is lowered, typical sizes of excitations are unchanged, but their concentration clearly decreases.  This concentration is recorded by
\begin{eqnarray}
\label{eqn:excon}
c_a  & = & 
 \Big \langle \cfrac{1}{V t_a }\, \sum_{i=1}^{N} h_i(0, t_a ; a) \Big \rangle  \nonumber \\ 
 & \approx & \Big \langle \cfrac{1}{V t_a  }\, \sum_{i=1}^{N} \theta \big(|\bar{\mathbf{r}}_i( t_a  + \langle \Delta t  \rangle_a )  - \bar{\mathbf{r}}_i(0)| - a \big) \Big \rangle \ , 
\end{eqnarray}
where the angle brackets denote equilibrium ensemble average, $V$ is the volume and $N$ is the total number of particles in the system.  This second (approximate) equality in Eq.~\ref{eqn:excon}, where $h_i(0, t_a ; a)$ is replaced by $\theta \left(|\bar{\mathbf{r}}_i( t_a  + \langle \Delta t  \rangle_a ) - \bar{\mathbf{r}}_i(0 )| - a \right)$, holds provided $t_a$ is indeed a plateau time.  At conditions where the liquid is not supercooled a separation of time scales need not exist, and thus there is no plateau behavior.  We replace $h_i$-functions with the $\theta$-functions when calculating $c_a$ at supercooled conditions.  

This quantity $c_a$ is the average number of displacements per unit space-time.  We refer to it as the average excitation concentration.  With transition path sampling, without the simplifying approximation of Eq.~\ref{eqn:excon}, we collect satisfactory equilibrium statistics for this average by accepting or rejecting trajectories with the weight functional formed from the indicator functional $h_i(t, t_a ; a)$ times the equilibrium distribution of trajectories, each trajectory having a duration of $t_a$.  This is done by first equilibrating the system with a trajectory that runs for many structural relaxation times.  A few short sections of this trajectory are then chosen to provide first examples of a transition or excitation associated with the tagged particle.  From these first trajectories, a series of shooting and shifting moves are then performed~\cite{bolhuis2002transition, dellago2002transtion} generating an ensemble of thousands of independent examples of excitations.  {The ensemble produced by this Monte Carlo walk through trajectory space is the subspace, with proper statistical weight, of those trajectories in the equilibrium distribution that exhibit excitation dynamics~\cite{bolhuis2002transition, dellago2002transtion}.} The commitment time $t_a$ is allowed to vary in transition path sampling~\cite{bolhuis2008rare} so that dynamics determines a distribution of these times without pre-conceived notions of its typical values.  In some cases, we have used transition path sampling to sample excitations in very cold systems out of equilibrium.  These systems are initialized from configurations of a warmer equilibrated system, but with temperatures chosen from a Maxwell-Boltzmann distribution at a lower temperature. 

Considering the range of possible displacement lengths, $a$ should be large enough to record significant displacements.  In addition, values of $a$ should be smaller than those for which typical trajectories would likely visit intermediate states for long periods of time.  Larger values of $a$ would obscure separations in time scales, and the equilibrium average $\langle h_i \rangle$ would fail to exhibit linear growth with respect to $t_a$~\cite{dellago1999calculation}.  For the systems that we consider here, taking $a$ to be 0.2 to 2 particle diameters proves to be satisfactory, and for this range of displacement lengths there is a range of suitable commitment times where $c_a$ is independent of $t_a$. At supercooled conditions, the typical mean values and fluctuations of $\Delta t$ and $t_a$ are of the order of $10^2$ to $10^3$ integration steps, and very much smaller than structural relaxation times. 

{Transition path sampling~\cite{bolhuis2002transition, dellago2002transtion} is a necessary tool in this study to extract information at very low temperatures.  At moderate supercooling conditions, however, where sufficient numbers of excitations are present at any one time frame, much of what we have done can be done with straightforward molecular dynamics.  Indeed, this is the procedure we use to evaluate $c_a$ from Eq.~\ref{eqn:excon}.  But in doing so, it is necessary to identify the duration of a time frame, $t_a$ or $\Delta t$.  Making that identification with transition path sampling is reasonably easy because the equilibrium distribution for instanton times can then be generated automatically~\cite{bolhuis2008rare}.  One can then limit the search for excitations with $\Delta t$ values not far from most probable values.  For a given displacement length $a$, especially for $a$ larger than a particle diameter, taking $\Delta t$ much larger than the most probable value will lead one to harvest sequences of several temporally separated excitations at smaller length scales.  With straightforward molecular dynamics, one can be assured of not using too large a value of $\Delta t$ by checking that this time is not beyond plateau value times, as discussed below.}

{Many features we highlight in this study of atomic glass forming liquids are also features of granular media~\cite{dauchot2005dynamical}.  In the context of granular media, these features have been studied with order parameters that focus on distinct changes in cages surrounding tagged particles~\cite{candelier2009building}.  Those order parameters might be useful alternatives to those considered in this work. }

\section{Localized excitations and hierarchical dynamics}

With the operational definition of excitation dynamics given above, the structure and energetics of this dynamics can be examined.  Thus, in this section we establish that the equilibrium statistics of excitation density is that of a dilute gas, and that the energetics of excitations grows logarithmically with the length scale of excitation displacement.  This logarithmic growth is the signature of a particular class of hierarchical dynamics, and we show that it predicts the temperature variation of structural relaxation times of a supercooled liquid.

\subsection{Dynamic indicators of excitations are localized with spatial and temporal extents that are temperature independent.}

\begin{figure}
\centerline{\includegraphics[width=1\columnwidth]{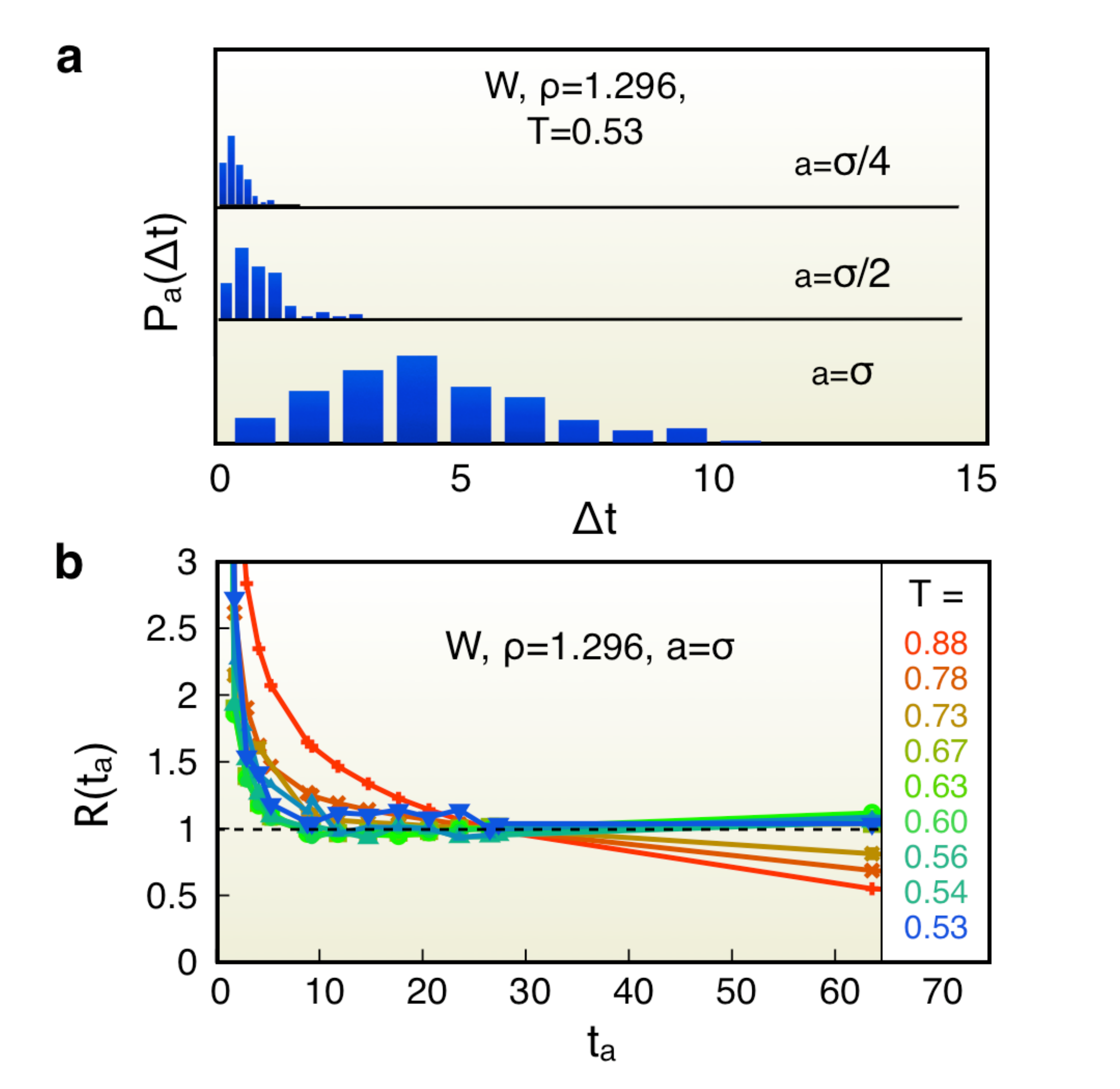}}
\caption{\label{fig:fig2} \textbf{Excitation dynamics exhibits separation of time scales with instanton times and plateau times}. ({\rm a}) The probability distribution of the instanton time, $P_a(\Delta t)$, for the W system (see Table~\ref{table:param}) at the lowest equilibrated temperature for three different displacement lengths, $a = \sigma/4$, $\sigma/2$ and $\sigma$. ({\rm b}) The mean value of the rate function $R(t_a)$ demonstrating that a plateau value exists for $T$ smaller than the onset temperature, $T_\textrm{o}$.  The particular system considered here is the W model at the density $\rho=1.296$, where $T_\textrm{o} \simeq 0.88$.  {Lines in Panel b are drawn through data points as guides to the eye.}}
\end{figure}

The time durations of {the dynamical processes that produce persistent displacements are} characterized by the probability distribution of the instanton time, $P_a(\Delta t)$.  The behavior of $P_a(\Delta t)$ for $a=\sigma$ is plotted in Fig.~\ref{fig:fig2}(a) for temperatures ranging from the onset temperature to the lowest  accessible temperature in our equilibrium molecular dynamics simulations.   The physical meaning of the onset temperature is seen explicitly in Fig.~\ref{fig:fig2}(b).  As $T$ goes below $T_\textrm{o}$, molecular motions are activated, and the concomitant separation of time scales is apparent from plateau values show in that figure.  Beyond the plateau region, {the function}
\begin{equation}
R(t_a) = \frac{\rho \langle \theta \left(|\bar{\mathbf{r}}_i(t_a + \langle \Delta t  \rangle_a) - \bar{\mathbf{r}}_i(0)| - a \right) \rangle}{ c_a t_a }
\end{equation}
will decrease with increasing $t_a$.  The range of the plateau region (i.e., the separation of time scales) increases with lowering $T$.  For $T$ close to or above $T_\textrm{o}$, however, motion is not activated and a plateau value does not exist.  (The determination of onset temperatures is presented in the next section).  {The existence of plateau behavior in $R(t_a)$ implies that statistically dominant motions are instantonic~\cite{chandler1978statistical,hanggi1990reaction}, like that illustrated in Fig.~\ref{fig:fig1}.}

\begin{figure*}
\centerline{\includegraphics[width=1\textwidth]{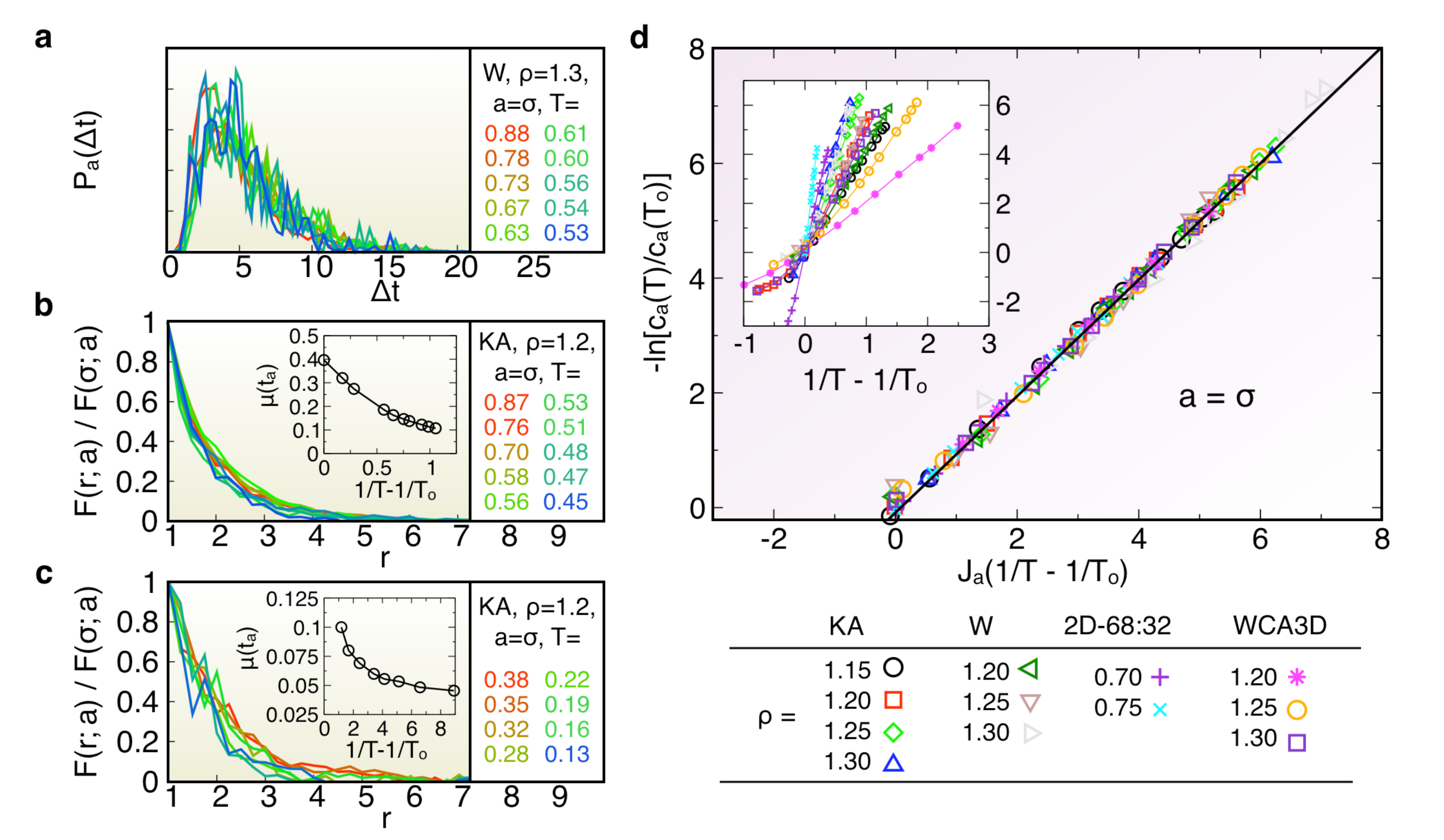}}
\caption{\label{fig:fig3} {\rm Excitations are localized and follow Boltzmann statistics of a dilute gas.}  ({\rm a}) Distribution of instanton times for excitations with displacement length $a=\sigma$ for the W model at several temperatures below its onset temperature, $T_\textrm{o} \simeq 0.88$.  ({\rm b})  Excess mobility density a distance $r$ from an excitation, relative to its value at $r=\sigma$ for a displacement length $a=\sigma$. The system considered here is the KA model at density $\rho=1.2$ at equilibrium conditions in the supercooled regime.  ({\rm c})  Same as (b) but for very cold non-equilibrium conditions, sampled using transition path sampling.  The insets of (b) and (c) show the corresponding temperature dependence of $\mu(t_a) = \left< \left| 
\bar{\mathbf{r}}_i(t_a) - \bar{\mathbf{r}}_i(0) \right| \right>$. ({\rm d}) Temperature dependence of excitation concentration $c_a$ for the 12 systems considered.  The inset shows plots of $-\ln c_a$ versus $1/T-1/T_\textrm{o}$ for excitations on a length scale $a=\sigma$.  The data collapse to the single curve when scaled by the energy parameter $J_a$.  Values of $T_\mathrm{o}$ and $J_a$ are given in Table~\ref{table:boltzmanndist}.}
\end{figure*}

Transport properties vary by three to four orders of magnitude over the range of temperatures considered in Fig.~\ref{fig:fig3}(a), but the distribution of instanton times varies little, and its width is far smaller than structural relaxation times at deeply supercooled temperatures (i.e., $T < T_\textrm{o}$).  In this sense, the extent of excitation dynamics in supercooled materials is relatively localized in time.  Changes in displacement length, shown in Fig.~\ref{fig:fig2}(a), produce significant changes in $P_a(\Delta t)$.  These changes relate to the hierarchical nature of the dynamics that we demonstrate in subsequent sections, where we also report on structural relaxation times.

Along with being temporally localized, excitation dynamics are spatially localized.  {To show this, we consider
\begin{eqnarray}
\label{eqn:mur}
\mu(r,t, t';a)  & = & \frac{1}{\langle h_1(0, t_a; a) \rangle} \nonumber \\
  &  & \times \, \Big \langle h_1(0, t_a; a)\,\sum_{i \neq 1}^{N} |\bar{\mathbf{r}}_i(t') - \bar{\mathbf{r}}_i(t)| \nonumber \\
 &   & \qquad \times   \, \delta(\bar{\mathbf{r}}_i(t) - \bar{\mathbf{r}}_1(t)- \mathbf{r})\Big \rangle ,
\end{eqnarray}
for $t=-t_a/2$ and $t' = t_a/2$, and with $t_a$ in the plateau regime.}  The function $\mu(r, t, t'; a)$ is the mean displacement density at $r$ for the time frame $t$ to $t'$ given an excitation is at the origin.  In the limit of large $r$, $\mu(r,t,t';a) \rightarrow \rho \mu(t-t')$, where $\rho$ is the mean particle density and $\mu(t-t') = \langle |\bar{\mathbf{r}}_i(t) - \bar{\mathbf{r}}_i(t')|\rangle$.  Deviations from this asymptotic limit reflect the degree with which particle displacements at $r$ are correlated to the excitation dynamics at the origin at time 0.  

This function has some similarity with the $\chi_4$-functions often used to to characterize dynamic heterogeneity~\cite{chi4, berthier2005direct, chandler2006lengthscale}.  Those functions, or susceptibilities, measure mean-square fluctuations of densities of particle-position overlap or displacement after a specified time frame.  For example, for the special case of taking $t =0$ and $t'-t = t_a$, the function $\mu(r,t,t';a)$ would be similar to the distinct particle contributions to a $\chi_4$-function at a time $t_a$.  But the $\mu$-function is different because it is not limited to that choice of time variables.  It is also different because it refers to particle coordinates that are coarse grained over time, while $\chi_4$-functions refer to raw coordinates.  Further, the $\mu$-function refers to a displacement density conditioned on an excitation at the origin, and an excitation is distinct from a displacement because an excitation produces a long-lived displacement and not simply a fleeting motion.  These features that distinguish the $\mu$-function from $\chi_4$-functions are important to the precision of our analysis.     

To use this function as a recorder of excitation size, the relevant time frame should surround time 0 and be of width $t_a$, so that a pertinent correlation function to consider is $\mu(r, -t_a / 2, t_a / 2; a) - \rho \mu(t_a)$.  This function oscillates as a function of $r$ manifesting local packing of molecules.  The same oscillations are found in the radial distribution function, $\langle \rho(\mathbf{r})\rangle_0$\,, where $\rho(\mathbf{r})$ is the net density in the time coarse-grained structure of particles at $\mathbf{r}$, and the average $\langle \cdots \rangle_0$ is taken with an additional particle fixed at the origin.  Throughout the supercooled regime, this radial distribution function is essentially temperature-independent, and its correlation length (the coarse-graining length over which its oscillations disappear) is of the order of one particle diameter.  Thus, for the purpose of viewing the spatial extent of a single excitation, it is useful to divide out the radial distribution function and consider
\begin{equation}
\label{eqn:fra}
F(r; a) = \frac{\mu(r, -t_a / 2, t_a / 2;a)}{ \langle \rho(\mathbf{r}) \rangle_0 \, \mu(t_a) }-1
\end{equation}
The quantity $F(r; a)$ is plotted relative to its value at $r=\sigma$ in Fig.~\ref{fig:fig3}(b).  The graphs show that, throughout the supercooled regime of the KA model, {motions in the same time frame correlate over only a small range of distances -- no more than a few atomic diameters -- and this correlation range  is independent of temperature.}  This observation holds for all equilibrium state points studied, Fig.~\ref{fig:fig3}(b), as well as  non-equilibrium state points sampled using transition path sampling, Fig.~\ref{fig:fig3}(c).  The same holds for all other models we have studied.

{The fact that motions in the same time frame correlate over only a small range of distances implies that the underlying configurations, which facilitate motion, must also be small with pair correlation lengths no larger than the motional correlation lengths seen in Panels (a) and (b) of Fig.~\ref{fig:fig3}.  A large underlying excitation or large correlation length between such excitations would imply large motional correlation lengths, and these are not observed.  In contrast, as we will see, much larger length scales that grow with decreasing temperature are associated with structural relaxation and dynamical heterogeneity.  In this sense, therefore, elementary excitations are spatially localized.  Growing length scales of structural relaxation must arise from correlations between those excitations at different time frames.}  These dynamical correlations are hierarchical, as we discuss next.

\begin{figure*}
\centerline{\includegraphics[width=0.95\textwidth]{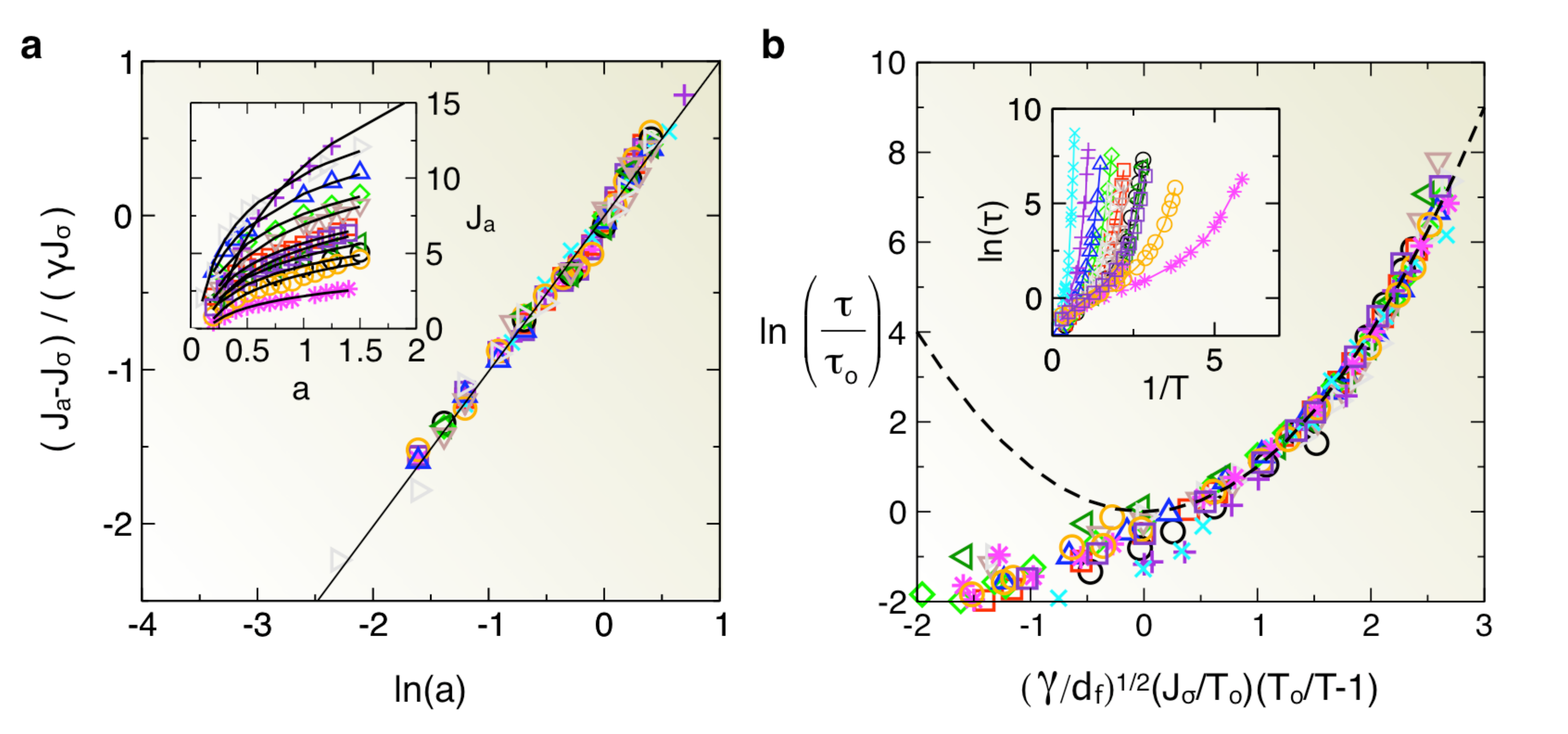}}
\caption{\label{fig:fig4} {\rm Hierarchical excitation energy scales predict observed structural relaxation times.}  ({\rm a}) Demonstration of logarithmic growth of $J_a$ with respect to $a$ (Eq.~\ref{eqn:Jloga}).  Fit parameters are given in Table~\ref{table:boltzmanndist}.  The inset shows the uncollapsed data.  The data set 2D-68:32, $\rho=0.75$ is omitted for clarity in the inset, as it scale obscures the other data sets.  ({\rm b}) Structural relaxation times predicted from results shown in Fig.~\ref{fig:fig3}d and Fig.~\ref{fig:fig4}a, which are compared to measured relaxation times.  Parameters from statistics are listed in Table~\ref{table:boltzmanndist}.  Data symbols are the same as those in Fig.~\ref{fig:fig3}.}
\end{figure*}

\subsection{Excitation densities obey Boltzmann statistics with energy scales that grow logarithmically with displacement length.}

We have computed the concentration of excitations $c_a$ from the second equality of Eq.~(\ref{eqn:excon}) and we find that $c_a$ obeys the Boltzmann distribution of Eq.~(\ref{eqn:cvsT}).  Figure~\ref{fig:fig3}(d) illustrates this finding.  For the length scale considered in that figure, $a=\sigma$, the plateau time $t_a$ is in the range $20$ to $30$.  The proportionality constant of Eq.~(\ref{eqn:cvsT}) is of the order of $(t_a a^d)^{-1}$, changing slightly from one system to another.  The onset temperature, $T_\textrm{o}$, signals the high temperature boundary for where $c_a$ is characterized by a single energy scale $J_a$.  We will see that this same temperature is the high-temperature end to all signatures of supercooled glassy dynamics.  The energy scale $J_a$ depends upon $a$ in a hierarchical way, Eq.~(\ref{eqn:Jloga}).  This finding is illustrated in Fig.~\ref{fig:fig4}(a).  The specific values of $J_\sigma$ and $\gamma$ that summarize these results are given in Table~\ref{table:boltzmanndist}.  

We provide two different columns of tabulated data for $J_\sigma$ because this quantity can be obtained by either of two ways -- by fitting data to Eq.~(\ref{eqn:cvsT}) or by fitting data to Eq.~(\ref{eqn:Jloga}).  It is significant that the two methods give similar values.  Those obtained in the former way have the better statistical certainty, so we use those values for predicting relaxation times, which we turn to now.

\begin{table*}
\caption{Parameters for data collapse in Figures~\ref{fig:fig3}d and~\ref{fig:fig4}a. }
\label{table:boltzmanndist}
\begin{tabular}{ | c | c  c  c  c  c || c  c  c c | }
\hline                        
Model & $\rho$ &  $J_{\sigma}$\footnotemark[1] & $ T_\textrm{o} $ & range\footnotemark[2] & $\Sigma$\footnotemark[3] & $\gamma$ & $J_{\sigma} \footnotemark[4] $ & $\tau_\textrm{o}$ & $\Sigma$\footnotemark[5] \\ \hline\hline
\  & 1.15 & $4.0\pm0.3$ & $0.67 \pm 0.03$ & 104 & 1e-3 & $0.55 \pm 0.05$ & $4.2 \pm 0.3$ & $90$ &1e-2 \\
KA & 1.2 & $5.3\pm0.2$ & $0.87 \pm 0.05$ & 111& 5e-4 & $0.41 \pm 0.04$ & $5.7 \pm 0.3$ & $120$ & 1e-2 \\
\  & 1.25 & $7.4 \pm 0.2$ & $1.06 \pm 0.05$ & 151 & 2e-4 & $0.39 \pm 0.03$ & $7.6 \pm 0.4$ & $160$ & 6e-3 \\
\  & 1.3 & $8.4\pm 0.3$ & $1.34 \pm 0.07$ & 271 & 5e-4 & $0.39 \pm 0.03$ & 9.0 $\pm 0.4$ & $130$ & 2e-3 \\
\hline
\  & 1.2 & $4.5  \pm 0.3$ & $0.66 \pm 0.05$ & 106 & 6e-4  & $0.59 \pm 0.1$ & $4.6 \pm 0.4$ & $240$ & 3e-2 \\
W & 1.25 & $6.9 \pm 0.2$ & $0.78 \pm 0.04$ & 92 & 6e-3 & $0.42 \pm 0.03$ & $6.9 \pm 0.4$ & $160$ & 1.2e-2 \\
\  & 1.296 & $9.4 \pm 0.3$ & $0.88 \pm 0.04$ & 227 & 2e-2 & $0.36 \pm 0.03$ & $9.3 \pm 0.3$ & $180$ & 9.6e-3 \\
\hline
2D-68:32 & 0.7 & $11.3 \pm 0.3$ & $1.36 \pm 0.05$ & 51 & 3e-3 & $0.62 \pm 0.05$ & $10.8 \pm 0.7$ & $65$ & 6.7e-3 \\
\  & 0.75 & $21.8 \pm 0.3$ & $2.13 \pm 0.1$ & 58 & 5e-3 & $0.57 \pm 0.03$ & $21.8 \pm 0.5$ & $60$ & 5e-3 \\
\hline
\  & 1.2  & 2.1 $\pm 0.1$ & $0.32 \pm 0.02$ & 57 & 2e-3 & $0.51 \pm 0.03$ & $2.2 \pm 0.2$ & $60$ & 1.2e-2 \\
WCA3D & 1.25 & $3.5 \pm 0.2$ & $0.45 \pm 0.03$ & 200 & 2e-3 & $0.51 \pm 0.03$ & $3.6 \pm 0.2$ & $135$ & 1.8e-2 \\
\  & 1.296 & $5.0 \pm 0.2$ & $0.58 \pm 0.03$ & 94 & 5e-4 & $0.47 \pm 0.03$ & $5.3 \pm 0.3$ & $120$ & 1.8e-2 \\ \hline
\end{tabular}
\begin{minipage}[t]{0.68 \textwidth}
\footnotetext[1]{From fits to Eq~\ref{eqn:cvsT} shown in Fig.~\ref{fig:fig3}d.}
\footnotetext[2]{The ratio of the maximum to minimum values of $c_a$ used in fits}
\footnotetext{to Eq~\ref{eqn:cvsT}.}
\footnotetext[3]{Error in the fit  to Eq~\ref{eqn:cvsT}, given by 1 minus the square of the}
\footnotetext{correlation coefficient.}
\footnotetext[4]{From fits to Eq~\ref{eqn:Jloga} shown in Fig.~\ref{fig:fig4}a.}
\footnotetext[5]{Error in the fit  to Eq~\ref{eqn:Jloga}, given by 1 minus the square of the}
\footnotetext{correlation coefficient.}
\end{minipage}
\end{table*}

\subsection{Relaxation times can be predicted from excitation energy scales.}

The logarithmic growth of $J_a$ with respect to $a$ has an important implication with respect to the temperature dependence of transport properties.  {In particular, and in contrast with excitations simply diffusing as a random walker, the logarithmic growth implies a hierarchical dynamics as imagined by Palmer et al.~\cite{palmer1984models} in which excitations on one scale combine to build excitations on larger scales. To see how this implies a specific temperature dependence of transport, consider excitations of displacement length $a$ for a time slice with thickness of order $t_a$.}  The spatial volume $V$ will be occupied by $Vc_a/ \nu $ of these excitation displacements, where $1/ \nu $ is of the order of $t_a a^d$.  The rate at which a given excitation will dissipate, $1/\tau_a$, is the rate at which that excitation can connect to neighboring excitations  of that same displacement $a$, and the energy to build that connection is its activation energy.  Therefore,
\begin{equation}
\label{Jell}
1/\tau_a \approx \nu\, \exp \left[-\left(J_{\ell_a} - J_a \right) \left(1/T - 1/T_\mathrm{o}\right) \right],\,\, T<T_\mathrm{o}\,
\end{equation}
where
\begin{equation}
\label{ella}
\ell_a/a = (c_a\,a^d)^{-1/d_\mathrm{f}}
\end{equation}
is the distance to connect a neighboring pair of excitations.  Motions in East-like models are nearly linear \cite{ritort2003glassy,garrahan2003coarse}, so that the fractal dimensionality, $d_\mathrm{f}$, is expected to be close to the physical dimensionality $d$.  It can be less than $d$ to the extent that paths connecting excitations are not linear.  Equations (\ref{Jell}) and (\ref{ella}) together with (\ref{eqn:Jloga}) yield
\begin{equation}
\label{taua}
\tau_a \,\nu = \exp\left[ J_a^2 \,(\gamma /d_\mathrm{f})\left(1/T - 1/T_\mathrm{o}   \right)^2 \right],\,\, T<T_\mathrm{o}.
\end{equation}
At $a \approx \sigma$, this relaxation time is the structural relaxation time for the systems we consider.  Hence, we find Eq.~(\ref{eqn:tauvsT}), with $J= J_\sigma \sqrt{\gamma/d_\mathrm{f}}\, $.  

To check this prediction we have determined structural times through calculations of the self correlation function
\begin{equation*}
F_s(k, t) = \langle \exp \{ i \mathbf{k} \cdot [ \textbf{r}_i(t) - \textbf{r}_i(0) ] \} \rangle .
\end{equation*}
For the three-dimensional models, we define the structural relaxation time, $\tau$, according to $1/e = F_s(q_0, \tau)$, where $q_0$ is the wave-vector at which the structure factor has its main peak $q_0 \approx 2\pi / \sigma$.  For the two-dimensional models, this definition is unsatisfactory because for those cases initial cage relaxation makes $F_s(q_0, t) < 1/e$ long before structural relaxation sets in; so for $d=2$, we use $0.1 = F_s(q_0, \tau)$.  Figure~\ref{fig:fig4}(b), shows the excellent agreement between measured structural relaxation times and theoretical prediction for all supercooled systems studied.  The reference structural relaxation time, $\tau_\textrm{o}$, is determined by computing a single relaxation time in the moderately supercooled regime and using this value to align the curves. { These values, listed in Table~\ref{table:boltzmanndist}, are approximations to the relaxation times of their respective liquids at the onset to supercooled behavior.}

The fractal dimension $d_\textrm{f}$ can be computed directly \cite{theiler1990estimating} by clustering neighboring particles that have displaced a distance $a=\sigma$ over a time window $\tau$.  The value of $d_\textrm{f}$ is then $d_\textrm{f}  = \mathrm{d} \ln(n) / \mathrm{d} \ln (r)$, where $n$ is the number of particles contained within a spherical volume of radius $r$ positioned at the cluster centroid.  We find that this procedure gives $d_\textrm{f}$ between 2.5 and 2.7 for physical dimension $d=3$, and between 1.8 and 1.9 for $d=2$.  Any values of  $d_\textrm{f}$ within those ranges are equally satisfactory for fitting the transport data.  The values of $d_\textrm{f}$ used in Fig.~\ref{fig:fig4}(b) are $d_\textrm{f} = 2.6$ for the three-dimensional systems and $d_\textrm{f} = 1.8$ for the two-dimensional systems, consistent with $0.8 d$ and $0.9 d$, respectively.  The one-dimensional version of the dynamical scaling we find is that of the East model~\cite{eastmodel}, where $d_\mathrm{f} = d = 1$.  {The remarkable data collapse shown in Fig.~\ref{fig:fig4}(b) is not the result of simple curve fitting.  As $J_\sigma$, $d_\mathrm{f}$, $T_\mathrm{o}$ and $\gamma$ are all determined independent of $\tau$, the collapse of data is a demonstration of the dynamical picture we have derived.}

\section{Dynamical facilitation, directionality, and stringlike motion}

The hierarchical nature of structural relaxation that we have uncovered in the previous two sections implies that dynamical facilitation~\cite{fredrickson1984kinetic,chandler2010dynamics} plays a central role in relaxation phenomena.  Dynamical facilitation refers to structural rearrangements or excitations allowing for the birth and death of excitations nearby in space.  The superposition of such dynamics leads to the formation of excitations on longer length and time scales, thus resulting in dynamical heterogeneity~\cite{garrahan2002geometrical}.  To illustrate these features more explicitly, we first consider computer rendered movies of trajectories, and then turn to quantitative analysis.

\subsection{Facilitation and hierarchical dynamics are evident in movies of supercooled liquid dynamics.}

\begin{figure*}
\begin{center}
\centerline{\includegraphics[width=0.85\textwidth]{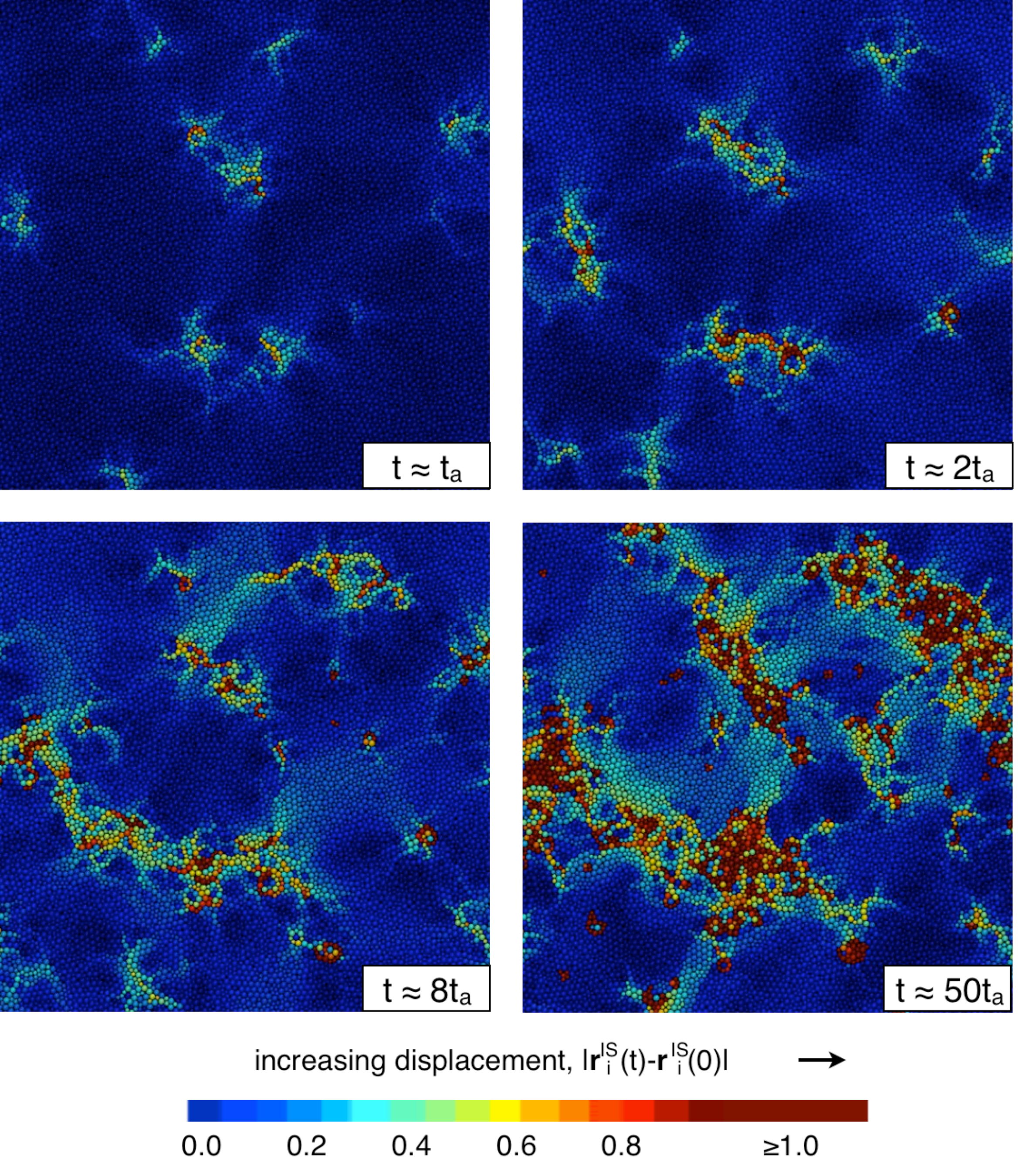}}
\caption{\label{fig:fig5} Click to view video clips: [\href{https://www.youtube.com/watch?v=SrP_5nOFi18}{Movie 1}] [\href{https://www.youtube.com/watch?v=01LaEvAdBUU}{Movie 2}].  Time evolution of particle displacements illustrate hierarchical dynamics in the 2D-50:50 system well below its onset temperature.  Specifically, this trajectory runs at a temperature $T=1.1$, and the onset temperature is $T_\mathrm{o} \approx 2.0$.  The movie depicts the displacement of each particle in terms of energy-minimized inherent structure coordinates $| \mathbf{r}^{\mathrm{IS}}_i(t) - \mathbf{r}^{\mathrm{IS}}_i(0) | $, with blue indicating overlap with initial position and red indicating a displacement of at least on particle diameter.  See key.   The movie, with frames separated by a reduced time of 25, which is approximately equal to $t_a$ for this system, spans a time of about 5 $\tau$, where the structural relaxation time is $\tau \approx 500 t_a$.  At very early times, non-oscillatory particle motions are sparse and spatially decorrelated, indicating the appearance of the initial excitations.  As time progresses, regions of high mobility (red, yellow) grow outward from the initial excitations until eventually, the regions connect on a timescale of about $\tau$.  Trajectories for the same system at temperatures above the onset temperature do not exhibit these features of correlated dynamics.}
\end{center}
\end{figure*}

An example of a trajectory is shown in Fig.~\ref{fig:fig5} for the 2D-50:50 system.  The movie depicts inherent structures, with particles colored according to the length of their displacement vector over a time window $t$.  Movies with time coarse-grained structures are similar, but we choose to show those of inherent structures to highlight the underlying physics of a most striking feature in these movies.  These features are  the ubiquitous low-frequency low-amplitude motions that permeate the system (manifested by aqua-colored particles).  These motions are not low frequency harmonic modes, as harmonic motions are absent from the inherent structure.  Rather, these motions are soft, anharmonic motions of the disordered system.  On time scales that are long compared to the period of these small amplitude soft motions, significant long-lived particle displacements occur (yellow, red particles).  These motions are associated with what appear to be defects or excitations in the system.  String-like patterns of slight mobility (aqua and green particles) surge outward from, and retract back towards, the initial excitations~\cite{chandler2010dynamics}.  

Eventually, rare, strong surges cause local regions to irreversibly deform on larger length scales and time scales.  These processes, whereby ubiquitous surging excitations play a significant role in producing rarer structural relaxations, are an intrinsic feature of East-like models.  Such surging was noted in Ref.~\cite{widmer2009central} but the connection to the East model was overlooked.  Activation energies grow as displacement length grows, so that frequencies of displacements decrease as length scales grow.  Further larger length scale displacements are built from smaller length scale displacements through facilitation.  This is why in the movie of Fig.~\ref{fig:fig5}, it is evident that short time local relaxation processes represent a microcosm of relaxation on longer length scales and time scales.  As time progresses, larger regions of mobility (red, yellow) grow outward from the initial relaxed regions, giving rise to more collective surges that are characterized by broader strings.  

\subsection{Dynamical facilitation, directionality and string-like motion are pronounced, and increasingly so as temperature is lowered.}

\begin{figure*}
\begin{center}
\includegraphics[width=0.95\textwidth]{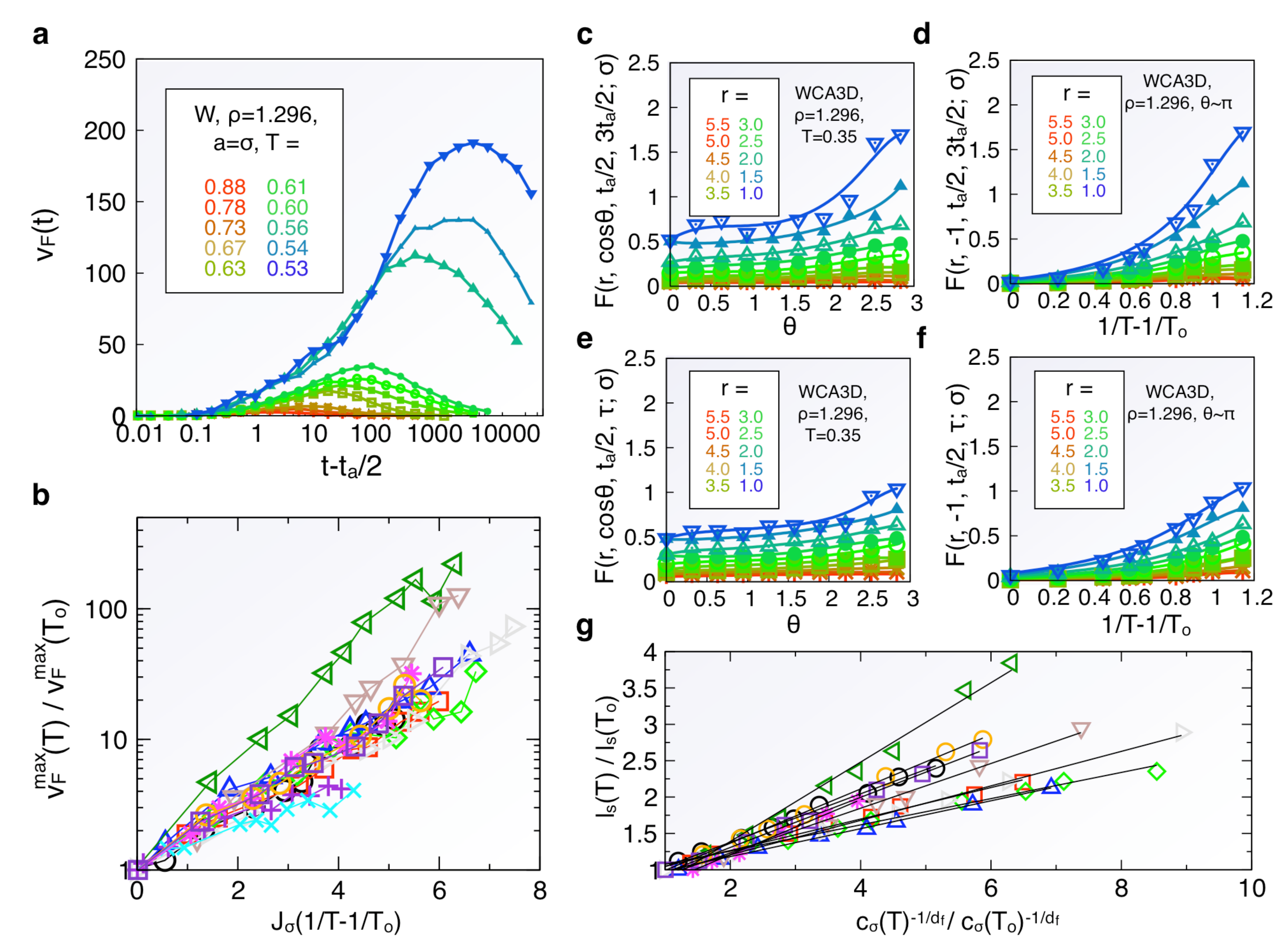}
\caption{\label{fig:fig6} {\rm Dynamics is facilitated, directional, and string-like, and more apparently so as temperature is lowered.}  (a) Dynamical facilitation volume, $v_\textrm{F}(t)$, for a representative W system at several different temperatures.  (b) The peak of the facilitation volume as a function of temperature for the 12 atomistic models studied.  (c) $F(r, \cos \theta, t, t'; a)$ for a single temperature with parameters $a=\sigma$, $t=t_a/2$, and $t'=3t_a / 2$ as a function of both the facilitation direction $\theta$ and the distance $r$ from the tagged particle at the origin at time $t_a/2$.  (d)  $F(r, \cos \theta, t, t'; a)$ as a function of the temperature with the same parameters as in (c) for a single facilitation angle $\theta \sim \pi$.  (e) $F(r, \cos \theta, t, t'; a)$ for a single temperature with parameters $a=\sigma$, $t=t_a/2$, and $t'=\tau$, where $\tau$ is the structural relaxation time, as a function of both the facilitation direction $\theta$ and the distance $r$ from the tagged particle at the origin at time $t_a/2$.  (f) $F(r, \cos \theta, t, t'; a)$ as a function of temperature with the same parameters as in (e) for a single facilitation angle $\theta \sim \pi$. (g) Mean string length $l_\mathrm{s}$ versus the concentration of excitations of displacement length $\sigma$, $c_\sigma$ at several different temperatures $T$ for the 10 three-dimensional systems studied.  Data symbols are the same as those in Fig.~\ref{fig:fig3}.}
\end{center}
\end{figure*}

While movies of particle motions are instructive, we obtain quantitative measures of hierarchical facilitation from the calculation of relaxation time presented in the previous section.  {We can also estimate} facilitation volumes,
\begin{equation}
\label{eqn:vfac}
v_\textrm{F}(t) = \int  \left[ \frac{ \mu (r, t_a / 2, t; a) } { \langle \rho(\textbf{r}) \rangle_0 \mu(t - t_a / 2) } -1 \right] d\mathbf{r}.
\end{equation}
The denominator dividing into the $\mu$-function in Eq.~(\ref{eqn:vfac}) is the value of the $\mu$-function in the absence of dynamical correlations with the initial excitation at the origin.  {We find that the integrand in Eq.\ref{eqn:vfac} looks much like the function $F(r;a)$ shown graphically in Fig.~\ref{fig:fig3}  b and c, but with a range that varies with time. }Thus, $v_\mathrm{F}(t)$ determines the volume of space where dynamics at time $t$ is correlated to that initial excitation.  In the absence of dynamical facilitation, particle displacements are not correlated with the initial excitation at the origin, and $v_\textrm{F}(t)$ goes to zero.  The {qualitative} behavior of $v_\textrm{F}(t)$ is shown in Fig.~\ref{fig:fig6}(a) for the W system at several different supercooled temperatures.  The other systems we have studied behave similarly.  

{At low temperatures, the sparsity of excitations makes it difficult to obtain accurate statistics for this facilitation volume.  Irregularities in the variations seen in Fig.~\ref{fig:fig6}(b) give a sense of the statistical uncertainties in our estimates for this quantity.  The data collected is sufficient to reveal the following trends:} The facilitation volume initially increases as a function of time $t$, as the initial excitations facilitate the formation of new excitations nearby.  $v_\textrm{F}(t)$ reaches a maximum value, $v^{\textrm{max}}_\textrm{F}(T)$, at $t \simeq \tau$, and this maximum value grows with decreasing temperature as the distance between the initial excitations increases.  Over time scales exceeding $\tau$, the relaxed regions overlap and the system becomes increasingly dynamically uniform with time, causing $v_\textrm{F}(t)$ to tend towards zero. Figure~\ref{fig:fig6}(b) shows that $v^{\textrm{max}}_\textrm{F}(T)$ consistently grows with decreasing temperature, thus demonstrating the presence of dynamical facilitation at all supercooled state points.

Dynamical facilitation occurs with directionality.  This is a feature of the $d=1$ East model~\cite{eastmodel} and its $d \geq 2$ counterparts~\cite{garrahan2003coarse,Berthier:2005hb}.  For the atomistic models studied herein, a quantitative measure of directionality in dynamical facilitation is given by~\cite{donati1998stringlike}
\begin{eqnarray}
\omega(r, \cos \theta, t, t';a) =  \quad \quad \quad  \quad \quad \quad \quad \quad \quad \quad \quad \quad \quad  \nonumber \\
\frac{1}{\langle h_1(0, t_a; a)\rangle }  \Big \langle h_1(0, t_a; a)\, \sum_{i \neq 1}^{N} |\bar{\mathbf{r}}_i(t') - \bar{\mathbf{r}}_i(t)|  \quad \\
\quad \quad \quad \quad  \times \, \delta(\cos \theta_i - \cos \theta)\, \delta(\bar{\mathbf{r}}_i(t) - \bar{\mathbf{r}}_1(t)- \mathbf{r})\Big \rangle , \nonumber 
\end{eqnarray}
where
\begin{equation}
\label{eqn:theta}
\cos \theta_i = \frac{ \bar{\textbf{r}}_i(t_a/2) \cdot[\bar{\mathbf{r}}_1(t_a / 2) - \bar{\mathbf{r}}_1 (-t_a / 2)|}{ | \bar{\textbf{r}}_i(t_a/2) | \,|\bar{\mathbf{r}}_1(t_a / 2) - \bar{\mathbf{r}}_1 (-t_a / 2 ) |}.
\end{equation}
Displacing particles that lie in the direction of  the tagged particle's displacement vector have $\theta_i \approx 0$, whereas particles that lie in the opposite direction have $\theta_i \approx \pi$.  

A measure of excess oriented motion is therefore~\cite{donati1998stringlike}
\begin{equation}
F(r, \cos \theta, t, t' ; a) = \frac{\omega(r, \cos \theta, t, t'; \sigma)}{ \langle \rho(\textbf{r}, \theta) \rangle_0 \mu(t' - t) } - 1,
\end{equation}
where 
$$
\langle \rho(\textbf{r}, \theta) \rangle_0 = \left< \sum_{i \neq 1}^{N} \delta(\cos \theta_i - \cos \theta)\, \delta(\bar{\mathbf{r}}_i(t) - \bar{\mathbf{r}}_1(t)- \mathbf{r}) \right>_0.
$$
Here, as before, $\langle \cdots \rangle_0$ denotes the average taken with particle 1 exhibiting the excitation at the origin.  Figure~\ref{fig:fig6}(c) shows that on relatively short time scales, excitations are most likely to occur behind earlier excitations a short distance away.  Figure~\ref{fig:fig6}(d) shows that this directional preference increases with decreasing temperature.  Figures~\ref{fig:fig6}(e)~and~\ref{fig:fig6}(f) show how directional effects dissipate with increasing time.

The latter observations are consistent with stringlike motion~\cite{donati1998stringlike, schroder1998hopping, glotzer2000spatially, aichele2003stringlike, gebremichael2004particle, vogel2004spatially, weeks2000three,keys2007measurement}.  On time scales of the order $t_a$, string-like motion takes the form of short ``microstrings,'' whose character does not change with the degree of supercooling~\cite{gebremichael2004particle}.  These motions are essentially synonymous with elementary excitations that we characterize here.  On timescales of the order $\tau$, microstrings combine to form longer strings, with a length scale that grows with the distance between the initial excitations.   In accordance with Refs.~\cite{donati1998stringlike},~\cite{ gebremichael2004particle}~and~\cite{keys2007measurement}, we construct strings by clustering particles that fully replace the initial position of neighboring particles over a time scale that maximizes the mean string length, $L_s$.  Strings constructed in this way have a minimum size of 2 particles; thus we subtract this baseline value from the average when quantifying changes in the mean string length, i.e., $l_\mathrm{s} = L_\mathrm{s} - 2$.  Figure~\ref{fig:fig6}(g) shows that $l_\mathrm{s}$ grows proportionally as a function of the mean distance between initial excitations, $c_a^{-1/d_\textrm{f}}$.  For our comparison, we consider excitations associated with particle displacements of length $a = \sigma$.  

Earlier ideas about dynamic heterogeneity, specifically the concepts of strings~\cite{donati1998stringlike, donati1999spatial, schroeder2000crossover, glotzer2000spatially, aichele2003stringlike, gebremichael2004particle, vogel2004spatially, keys2007measurement} and micro-strings~\cite{gebremichael2004particle}, seem to be connected with our results.  In particular, these and other earlier works highlighted the presence of correlated stringlike motions in the heterogeneous dynamics.  A small subset of these motions involve a few particles that move nearly simultaneously, a process which is called a microstring~\cite{gebremichael2004particle}.  Longer strings and then clusters of strings are built from these microstrings, processes that take place over times that are significantly longer than those of a microstring~\cite{gebremichael2004particle}.  In the context of this paper, microstrings coincide with dynamics of elementary excitations, and the building of longer strings and clusters of strings coincide with the facilitated hierarchical dynamics of excitations.  The exponential distribution of string lengths reported by earlier studies of dynamic heterogeneity in many types of liquids and in granular materials  ~\cite{donati1998stringlike, donati1999spatial, aichele2003stringlike, gebremichael2004particle, vogel2004spatially, keys2007measurement} may now be understood as a consequence of the ideal gas statistics obeyed by the elementary excitations.  Consequently, the cooperative length scale of stringlike motion that grows with decreasing temperature coincides with a growing distance between excitations.  It is a plausible connection that is worthy of further study.

\section{Discussion}

This paper shows that excitations are localized, with static correlations like those of an ideal gas, and that dynamics in these systems are non-trivial because motions occur in a fashion that is facilitated and hierarchical.  Perhaps most importantly, the number of particles involved in a single elementary excitation is independent of temperature.  As a result,  dynamics slow and dynamical heterogeneity length scales grow with decreasing temperature, not because the fundamental mechanism of particle motion becomes increasingly cooperative, but rather because the distance between excitations increases.  These findings would thus seem to conflict with pictures, such as the mosaic picture, based on ``cooperatively rearranging regions'' (CRRs) in which the regions themselves are considered the fundamental object. Instead, here, we view cooperativity as emerging hierarchically over time through facilitation, with excitations being the fundamental object.  As noted in the Introduction, we do not discount the possibility of a yet-to-be discovered link that would show how growing CRR size is related to the mean distances connecting neighboring excitations, $\ell_a \propto c_a^{-1/d_\mathrm{f}}$.  However, this would presumably require a reformulation of the definition of a CRR. The connection between string length and mean excitation distance demonstrated here might serve as a starting point~\cite{gebremichael2005spatially}.  While we can await that development, at this stage we conclude this paper by addressing the most common criticisms of the picture supported by our results.

Perhaps the foremost criticism is that no viable procedure has existed from which localized excitations could be seen as emergent properties of atomistic dynamics~\cite{biroli2009random, berthier2011theoretical}.  This paper has established such a procedure.  

Another criticism concerns the measured behavior of heat capacity.   Unlike reversible heat capacity in an equilibrium system, this property exhibits significant hysteresis near the glass transition temperature, $T_\mathrm{g}$.  Heat capacity per molecule at temperatures above the range of hysteresis is higher than that at temperatures below the range of hysteresis by an amount typically larger than Boltzmann's constant.  In some theories, including Adam and Gibbs', this change in heat capacity, $\Delta C$, is of central importance because it can be related to an imagined vanishing of configurational entropy at a finite temperature, and this vanishing is supposed to signal loss of ergodicity.  Inspection of experimental data discredits perceived correlations between relaxation times and these thermodynamic properties (see Fig.~1 of Ref.~\cite{elmatad2010corresponding}).  Nevertheless, interest in $\Delta C$ persists, and one may wonder whether its typical values are consistent with dynamics in structural glass formers pictured in terms of localized excitations~\cite{biroli2005defect}.

Because localized excitations emerge from molecular dynamics, it is unlikely that this picture is inconsistent with thermodynamics, at least for the systems studied herein.  Still, typical values of $\Delta C$ reflect the number of translational degrees of freedom removed upon cooling below the glass transition, and this number seems physically interesting.  The removal of translational freedom coincides with the removal of {fluctuations in the number of excitations}.  Therefore, a simple connection between $\Delta C$ and excitation concentration is $\Delta C \propto c_\sigma(T_\mathrm{g})$. This proportional relationship was proposed in Ref.~\cite{garrahan2003coarse}, where it was shown to be consistent with experimental data. The constant of proportionality must grow with the number of particles correlated with an excitation of unit displacement length.  We see from this paper that this number is of order 10 or more.  Whether further steps can be taken to predict precise values of the proportionality constant remains to be seen.  {These steps may involve attempts to understand connections between local vibrational modes and excitations.  The former are not causative of dynamics~\cite{ashton2009relationship}, but they are correlated to the latter~\cite{widmer2008irreversible}, and the former extends over regions of space that are large compared to the latter~\cite{ashton2009relationship}.}

{Another criticism is the recent suggestion that facilitation is not the cause of intermittency in dynamical activity (births of so-called ``avalanches'')~\cite{candelier2009building, candelier2010spatiotemporal}, and further that facilitation diminishes as temperature is significantly lowered below (or density is increased  above) the onset to supercooled behavior~\cite{candelier2010dynamical}.  These seemingly contradictory conclusions to our findings are} the result of differences in the definition of facilitation.  In our usage, facilitated dynamics is where changes in configurations or microstates occur only in the vicinity of excitations.  Excitations refer to configurations or microstates.  Thus, in facilitated dynamics, the birth or death of excitations occurs only in the presence of neighboring excitations.  From this definition, it follows that connected lines of excitations permeate trajectory space in systems dominated by facilitated dynamics~\cite{garrahan2002geometrical}.  

Although excitations will connect throughout space-time, changes in states will not necessarily connect in facilitated dynamics.  These changes are the so-called ``kinks'' of kinetically constrained models.  Kinks necessarily record the presence of one or more excitations, but excitations can persist for long periods without the appearance of kinks.  References~\cite{candelier2010spatiotemporal} and \cite{candelier2010dynamical} ascribe a loss of facilitation to growing spatial and temporal separations between changes in state, but these are growing separations between kinks, not excitations.  Indeed, the behaviors noted in those papers are predicted from the facilitated dynamics of the East model and its generalizations.  Specifically, the motion of particles in the presence of excitation lines is intermittent~\cite{jung2004excitation}.  Clustered bursts of activity occur at space-time regions where a particle or group of particles intersects excitation lines.  Bursts end and disconnect from other bursts when this intersection ends, after which the particle persists in an inactive state for a relatively long period.  This behavior is the essential element of decoupling phenomena~\cite{hedges2007decoupling}, and it becomes more striking as relaxation times grow because the lengths of these quiescent periods grow with increased supercooling\cite{jung2004excitation, hedges2007decoupling}.   Whether this understanding can be enhanced from analysis presented in Refs.~\cite{candelier2010spatiotemporal}~and~\cite{candelier2010dynamical} remains to be seen.

\begin{acknowledgments}
The National Science Foundation supported ASK, LOH, SCG and DC in the development of computational tools implementing transition path sampling methods under Grant No. CHE-0624807.  DC and ASK were supported in the final stages by DOE Contract No. DE-AC02Ñ05CH11231.  LOH performed portions of this work as a User project at the Molecular Foundry, Lawrence Berkeley National Laboratory, which is supported by the Office of Science, Office of Basic Energy Sciences, of the U.S. Department of Energy under Contract No. DE-AC02Ñ05CH11231.  We thank T. Speck, U.R. Pedersen, and Y.S. Elmatad for helpful discussions.  We thank D.T. Limmer and P. Varilly for helpful comments regarding the manuscript.
\end{acknowledgments}


%

\end{document}